\documentstyle[12pt]{article}

\input epsf
\begin{document}
\begin{titlepage}
\today          \hfill
\begin{center}
\hfill    OITS-687\\

\vskip .05in

{\large \bf
Penguins and Mixing Dependent CP Violation
}
\footnote{This work is supported by DOE Grant DE-FG03-96ER40969.}
\footnote{To appear in Proceedings of the Third International Conference on
B Physics and CP Violation, Taipei, December 3-7, 1999, H.-Y. Cheng and 
W.-S. Hou, eds.
(World Scientific, 2000).}
\vskip .15in
N.G. Deshpande
\footnote{email: desh@oregon.uoregon.edu}
\vskip .1in
{\em
Institute of Theoretical Science \\
University
of Oregon \\
Eugene OR 97403-5203}
\end{center}

\vskip .05in

\begin{abstract}
Constraints on angles of Unitarity triangle are reviewed, and
in particular constraint on $\gamma$ from limit on $\Delta m_s$ is
emphasized.  Effects of penguin diagram on measurement of $\beta$
and $\alpha$ are then reviewed.  New measurements on $B \rightarrow
\pi^+\pi^-$ in QCD improved factorization approximation suggest
large penguin effects.  It is possible to estimate the error in
measurement of $\alpha$ as a function of $\gamma$ for different
$\left|V_{ub} /V_{cb}\right|$ values.  

\end{abstract}

\end{titlepage}

\newpage
\renewcommand{\thepage}{\arabic{page}}
\setcounter{page}{1}

\section{Constraints on Angles of Unitarity Triangle}

Constraints on unitarity triangle expressed in terms of Wolfenstin
parameterization
are given below:

\begin{description}
\item[[a]]  From charmless semileptonic B decays \cite{pdg}:
\begin{eqnarray}
\left|{V_{ub} \over V_{cb}}\right| = 0.08 \pm 0.02
\label{a1}
\end{eqnarray}
which yields
\begin{eqnarray}
\left(\rho ^2 + \eta ^2\right)^{1/2} = 0.36 \pm 0.09
\label{a2}
\end{eqnarray}

\item[[b]] From $B_d - \bar{B}_d$ mixing.  Error is dominated by $f_B$ and the
bag factor $B_B$:
\begin{eqnarray}
\left|V_{td}\right| = 0.009 \pm 0.003
\label{b1}
\end{eqnarray}
which yields
\begin{eqnarray}
\left|1 - \rho- i \eta \right| = 1.0 \pm 0.3
\label{b2}
\end{eqnarray}

\item[[c]] Value of $\epsilon$ in $K$ system.
Error is dominated by hadronic matrix elements:
\begin{eqnarray}
\eta \left(1 - \rho + 0.35\right) = 0.48 \pm 0.20
\label{c1}
\end{eqnarray}

\item[[d]] Constraints from $B_s - \bar{B}_s$ \cite{ms}
\begin{eqnarray}
\left(\Delta m_s\right) > 14.3 ps^{-1} \left(90\% CL\right)
\label{d1}
\end{eqnarray}
Using the relation from Box diagrams
\begin{eqnarray}
\left|{V_{td}  \over  V_{ts}}\right| = \xi    \left[{m_{B_s} \Delta m_d
\over  m_{B_d} \Delta m_s}
\right]^{1/2}
\label{d2}
\end{eqnarray}
where
\begin{eqnarray}
\xi = {f_{Bs}  \over  f_{Bd}} \sqrt{{B_{Bs}  \over  B_{Bd}}}
\label{d3}
\end{eqnarray}
Lattice calculations yield for $\xi$ the value \cite{lattice}
\begin{eqnarray}
\xi = 1.15 \pm 0.05
\label{d4}
\end{eqnarray}
This translates into the bound
\begin{eqnarray}
\left|V_{td} / V_{ts}\right| < 0.214
\label{d5}
\end{eqnarray}
or
\begin{eqnarray}
\left|1 - \rho- i \eta \right| < 0.96
\label{d6}
\end{eqnarray}
\end{description}

The last constraint in particular
implies
$\gamma < 90^{\circ}$ and $75^{\circ} < \alpha < 120^{\circ}$.  This allowed
range of $\gamma$ 
leads to unique estimate of errors in ${\alpha}$ as we shall
see.

\section{CP Violation Through Mixing}

Strategy to measure $\beta$ and $\alpha$ involve measuring time dependent
asymmetry in $B$ decays to CP eigenstate.  Defining the long and short lived
eigenstates of $B$ as
\begin{eqnarray}
\left| \left. B\right. \right>_{L,S} = p\left| \left. B^{\circ}\right.
\right> \pm q
\left| \left. \bar{B}^{\circ}\right. \right>,
\label{II1}
\end{eqnarray}
the amplitudes for decays into CP eigenstates are defined as
\begin{eqnarray}
A = \left< f_{CP} | H_w | B^{\circ}\right>
\label{II2}
\end{eqnarray}
\begin{eqnarray}
\bar{A} = \left< f_{CP} | H_w | \bar{B}^{\circ}\right>.
\label{II3}
\end{eqnarray}
The asymmetry is then defined by
\begin{eqnarray}
Asy (t) = \left[\left(1 - \left|\lambda\right|^2\right) \cos
\left(\Delta Mt\right)- \ 2 Im (\lambda) \sin \left(\Delta Mt\right)
\right] /
\left[| 1 + |\lambda |^2 \right]
\label{II4}
\end{eqnarray}
where $\lambda = (q/p)\left(\bar{A}/A\right)$.  In the standard model
$(q/p) = e
^{-2i\beta}$.  If $A$ is dependent on a single weak
phase,
\begin{eqnarray}
\left(\bar{A}/A\right) = e^{-2i\phi_{\rm{weak}}}
\label{II5}
\end{eqnarray}
then we have the expression
\begin{eqnarray}
Asy (t) = - Im (\lambda) \sin \left(\Delta Mt\right)
\label{II6}
\end{eqnarray}

\subsection{Measurement of $\beta$}
The mode that has the least theoretical uncertainty is $B \rightarrow \psi
Ks$.
The amplitude for this mode can be written in terms of Tree and Penguin
contribution as
\begin{eqnarray}
A = V_{cb}V^{\ast}_{cs} T + V_{tb}V^{\ast}_{ts} P  =
V_{cb}V^{\ast}_{cs} (T-P) + V_{ub}V^{\ast}_{us} P
\label{II7}
\end{eqnarray}
since $\left|V_{ub}V^{\ast}_{us}/V_{cb}V^{\ast}_{cs}\right|$
 is $\approx 1/50$, and the Penguin contribution has predominantly
$\bar{c}c$ in a color octet state, the contribution due to penguin
diagram is less than 1\%.

If $B \rightarrow D^+D^-$ mode is used instead, the penguin
contribution is much larger, and there is no color suppression either.
\begin{eqnarray}
A = V_{cb}V^{\ast}_{cd} T + V_{tb}V^{\ast}_{td} P  =
V_{cb}V^{\ast}_{cd} (T-P) - V_{ub}V^{\ast}_{ud} P
\label{II8}
\end{eqnarray}
The value of $\left| V_{ub}V^{\ast}_{ud}/V_{cb}V^{\ast}_{cd}\right|
\approx 0.3$, and although $P$ is suppressed compared to $T$
due to small Wilson coefficients, one
can expect a contamination due to penguin of a few percent.

\subsection{Measurement of $\alpha$}

The mode $B^{\circ} \rightarrow \pi^+\pi^-$ lends itself to the
earliest measurement of $\alpha$.  For this mode the amplitude is
\begin{eqnarray}
A = V_{ub}V^{\ast}_{ud} T + V_{tb}V^{\ast}_{td} P =
V_{ub}V^{\ast}_{ud} (T-P) + V_{cb}V^{\ast}_{cd} P
\label{II9}
\end{eqnarray}
The value of $\left| V_{cb}V^{\ast}_{cd}/V_{ub}V^{\ast}_{ud}\right|
\approx 3$ giving a crude estimate of around 15\% for the penguin
contamination.  Gronau and London \cite{gl} have presented a method of
extracting $\alpha$ from measurements of $B^{\circ} \rightarrow
\pi^+\pi^-$, $\overline{B^{\circ}} \rightarrow \pi^+\pi^-$,
$B^{\circ} \rightarrow \pi^{\circ}\pi^{\circ}$, $\overline{B^{\circ}}
\rightarrow
\pi^{\circ}\pi^{\circ}$ and $B^+ \rightarrow \pi^+\pi^{\circ}$.
However, the most recent theoretical estimates of $B^{\circ}
\rightarrow \pi^{\circ}\pi^{\circ}$ branching ratio are around $ 5
\times 10^{-7}$, making this method academic at present.  However,
we now discuss theoretical developments that may allow us to extract
the correct $\alpha$ from measurements of asymmetry in
$B^{\circ} \rightarrow \pi^+\pi^-$ alone.

\section{Determination of $\alpha$ from $B^{\circ} \rightarrow
\pi^+\pi^-$}
\begin{figure}
\vspace{-0.4in}
\centerline{\epsfxsize=0.8\textwidth \epsfbox{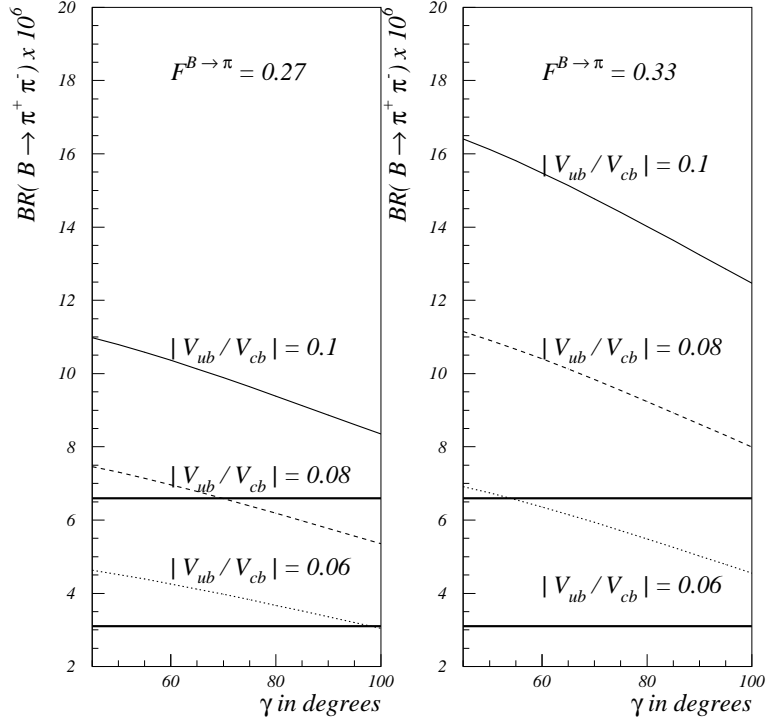}}
\vspace{-0.25in}
\caption{$CP$-averaged $BR \left( B \rightarrow \pi^+ \pi^- \right)$
as a function of $\gamma$ for $F^{B \rightarrow \pi^-} = 0.27$
(left) and $0.33$ (right)
and
for
$| V_{ub} / V_{cb} | = 0.1$
(solid curves), $0.08$ (dashed curves) and $0.06$
(dotted curves). The BR measured by the CLEO
collaboration lies (at the $1 \; \sigma$ level)
between the two horizontal (thicker) solid lines. The errors on the
CLEO measurement have been added in quadrature to compute the $1 \;
\sigma$  limits.}
\protect\label{figpi+pi-}
\protect\label{figpi+pi-1}
\end{figure}

This is based on recent work of Agashe and Deshpande \cite{ad}.
Recently, the CLEO collaboration has reported the first observation of the
decay
$B \rightarrow \pi^+ \pi^-$ 
\cite{poling}.The effective Hamiltonian for $B$ decays is:
\begin{eqnarray}
{\cal H}_{eff} & = & \frac{G_F}{\sqrt{2}}
\Biggl[
V_{ub} V_{ud}^{\ast}
\left( C_1 O_1^u + C_2 O_2^u \right)
\Biggr.
\nonumber \\
& & \Biggl. + V_{cb} V_{cd}^{\ast} \left( C_1 O_1^c + C_2 O_2^c \right) -
V_{tb} V_{td}^{\ast} \sum _{i=3} ^6 C_i O_i
\Biggr].
\end{eqnarray}
The $C_i$'s are the Wilson coefficients (WC's).
In a recent paper, Beneke {\it et al.} found that the matrix elements
for the decays $B \rightarrow \pi \pi$, in the large
$m_b$ limit, can be written as \cite{beneke}
\begin{eqnarray}
\langle \pi \pi | O_i | B \rangle & = &
\langle \pi | j_1 | B \rangle \langle \pi | j_2 | 0 \rangle \nonumber \\
& & \times \Bigr[ 1 + \sum r_n \alpha _s^n (m_b) + O ( \Lambda _{QCD}
/ m_b ) \Bigl],
\end{eqnarray}
where $j_1$ and $j_2$ are bilinear quark currents.
If the radiative corrections in $\alpha _s$ {\em and}
$O ( \Lambda _{QCD}
/ m_b )$ corrections are neglected, then the
matrix element on the left-hand side
factorizes into a product of a form factor and a meson decay constant
so that we recover the ``conventional'' factorization formula.
These authors computed the $O(\alpha _s)$ corrections.
In this approach, the strong interaction (final-state rescattering)
phases are included in the radiative corrections in $\alpha _s$
and thus the $O(\alpha _s)$ strong interaction
phases are determined \cite{beneke}.
The matrix element for $B \rightarrow \pi^+ \pi^-$ is \cite{beneke}:
\begin{eqnarray}
i \bar{A}\left( \bar{B}_d \rightarrow \pi^+
\pi^- \right)
& = & \frac{G_F}{\sqrt{2}} \Biggl[ V_{ub} V_{ud}^{\ast} \left(
a_1 + a_4^u +
a_6^u r_{\chi} \right) \Biggr. \nonumber \\
& & + \Biggl.
V_{cb} V_{cd}^{\ast} \left( a_4^c +
a_6^c r_{\chi} \right) \Biggr] \times X.
\label{bpi+pi-}
\end{eqnarray}
Here
\begin{equation}
X = f_{\pi} \left( m_B^2 - m_{\pi}^2 \right)
F_0^{B \rightarrow \pi^-} \left( m_{\pi}^2 \right),
\label{X}
\end{equation}
where $f_{\pi} = 131$ MeV is the pion decay
constant and $F_0^{B \rightarrow \pi^-}$ is a 
form factor.
In the above equations, the $a_i$'s are (combinations of) WC's with the
$O(\alpha _s)$ corrections added.  The values of the $a_i$'s are given in
Table \ref{a} \cite{beneke}. The imaginary parts of $a_i$'s are due to
final-state rescattering. For the $CP$ conjugate processes, the CKM elements
have to be
complex-conjugated.
We discuss two values of the form factors: $F^{B \rightarrow \pi^-}
= 0.27$ and $0.33$.
Model calculations indicate that
the $SU(3)$ breaking in the form factors is given by
$F^{B \rightarrow K^-} \approx 1.13 \; F^{B \rightarrow \pi^-}$
\cite{bsw,lightcone}.
The large measured
$BR( B \rightarrow K \eta^{\prime} )$
requires $F^{B \rightarrow K^-} \stackrel{>}{\sim} 0.36$
\cite{desh} which, in turn, implies a larger value of
$F^{B \rightarrow \pi^-}$ $(\approx 0.33)$. If $F^{B \rightarrow K^-}
\stackrel{<}{\sim} 0.36$,
then we require a ``new'' mechanism to account for
$BR( B \rightarrow K \eta^{\prime} )$: high charm content of
$\eta^{\prime}$ \cite{ali}, QCD anomaly \cite{qcd}
or new physics. Also, if
$F^{B \rightarrow \pi^-} < 0.27$, then the
value of $F^{B \rightarrow K}$ is too small to explain the measured
BR's for $B \rightarrow K \pi$ \cite{dutta}.
We use
$| V_{cb} | = 0.0395$, $|V_{ud}| =0.974$,
$|V_{cd} | = 0.224$, $m_B = 5.28$ GeV and $\tau _B = 1.6$ ps \cite{pdg}.
\renewcommand{\arraystretch}{1}
\begin{table}
\begin{center}
\begin{tabular}{|c|c|} \hline
$a_1$ & $1.047 + 0.033 \; i$\\ \hline
$a_2$ & $0.061 - 0.106 \; i$\\ \hline
$a_4^u$ & $-0.030 - 0.019 \; i$\\ \hline
$a_4^c$ & $-0.038 - 0.009 \; i$\\ \hline
$a_6^{u,c} \; r_{\chi} $ & $-0.036$\\ \hline
\end{tabular}
\end{center}
\caption{The factorization coefficients for the renormalization
scale $\mu = m_b /2$ \protect\cite{beneke}.}
\label{a}
\end{table}
In
Fig. \ref{figpi+pi-}
we show the $CP$-averaged BR for $B \rightarrow \pi^+ \pi^-$
as a functions of
$\gamma$ for $F^{B \rightarrow \pi^-} = 0.33$ and $0.27$ and
for $| V_{ub} / V_{cb} | = 0.1$, $0.08$ and $0.06$.

The CLEO measurement is $B \rightarrow \pi^+ \pi^- = \left(
4.7^{+1.8}_{-1.5}
\pm 0.6 \right) \times 10^{-6}$ \cite{poling}.
If
$F^{B \rightarrow \pi^-} = 0.33$
and for $\gamma \stackrel{<}
{\sim} 90^{\circ}$,
we see from the figures that
smaller values of $| V_{ub} / V_{cb} |
\approx 0.06$ are preferred: $|V_{ub} / V_{cb}|
=0.08$
is still allowed at
the $2 \sigma$ level for $\gamma \sim 100^{\circ}$.
The smaller value of $| V_{ub} / V_{cb} |$ leads to
greater penguin contamination.
However, if the smaller value of the form factor ($0.27$) is used, then
the CLEO measurement is consistent with $|V_{ub} / V_{cb}|
\approx 0.08$.
We obtain similar results using
``effective'' WC's ($C^{eff}$)'s and $N = 3$ in the earlier
factorization framework
\cite{ali}.

\begin{figure}
\vspace{-0.5in}
\centerline{\epsfxsize=0.8\textwidth \epsfbox{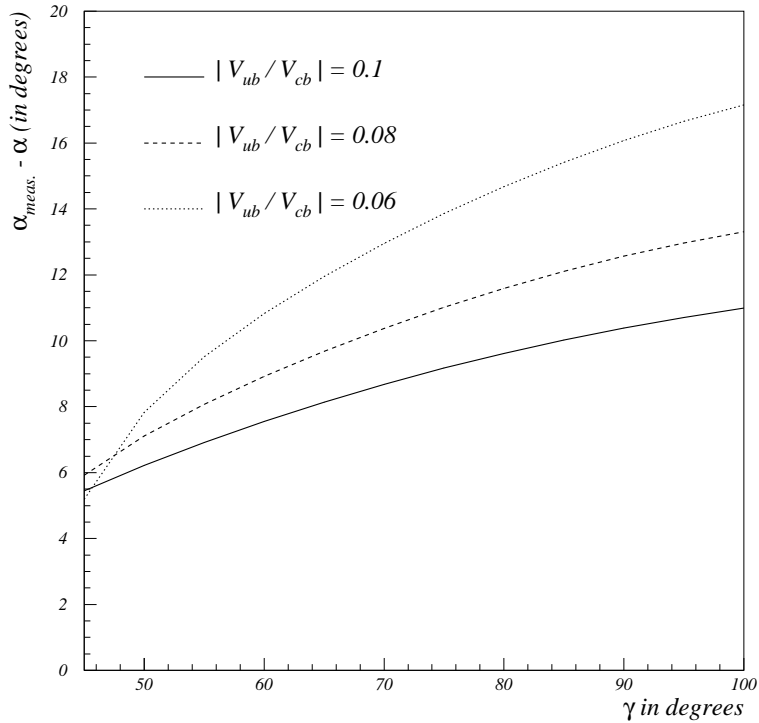}}
\vspace{-0.25in}
\caption{The error in the measurement of CKM phase $\alpha$
using (only) time-dependent $B \rightarrow \pi^+ \pi^-$ decays as a
function of
$\gamma$
for
$| V_{ub} / V_{cb} | = 0.1$ (solid curve), $0.08$ (dashed curve) and
$0.06$ (dotted curve).}
\protect\label{figealpha}
\end{figure}

Since the $B_d-\bar{B}_d$ mixing phase is $2 \beta$,
 if we neglect the
(QCD) penguin operators,
{\it i.e.}, set
$a_{4,6} = 0$ in Eq. (\ref{bpi+pi-}), we
get
\begin{equation}
\frac{\bar{A}}{A} = e^{- i 2 \gamma}
\end{equation}
and
\begin{equation}
\hbox{Im} \lambda = \sin \left( - 2 (\beta + \gamma) \right) = \sin 2
\alpha.
\label{alphanop}
\end{equation}
In the presence of the penguin contribution, however,
$\bar{A}/{A} \neq e^{-i 2 \gamma}$ so that $\hbox{Im} \lambda
\neq
\sin 2 \alpha$. We define
\begin{equation}
\hbox{Im} \lambda = \hbox{Im} \left( e^{-i2 \beta} \frac{\bar{A}}
{{A}}
\right) \equiv
\sin 2 \alpha_{meas.}
\label{alphameas}
\end{equation}
as the ``measured'' value of $\sin 2 \alpha$,
{\it i.e.}, $\sin 2 \alpha _{meas.} = \sin 2 \alpha$ if the penguin
operators
can be neglected.
In
Fig. \ref{figealpha}
we plot the error in the measurement of
$\alpha$, $\Delta \alpha \equiv \alpha_{meas.} - \alpha$, where
$\alpha_{meas.}$ is obtained
from Eq. (\ref{alphameas}) 
and
$\alpha$ is obtained from $\gamma$ and $| V_{ub} / V_{cb} |$.
Note that
$\Delta \alpha$
is independent of
$F^{B \rightarrow \pi^-}$ since the form factor
cancels in the ratio $\bar{A} / {A}$.
We see that for the values of
$| V_{ub} / V_{cb} |
\approx 0.06$ preferred by the $B \rightarrow \pi^+ \pi^-$ measurement
(if $F^{B \rightarrow \pi^-} \approx 0.33$),
the error in the determination of $\alpha$ is large $\sim 15^{\circ}$ (for
$\gamma \sim 90^{\circ}$). If $F^{B \rightarrow \pi^-} \approx 0.27$, then
$| V_{ub} / V_{cb} |
\approx 0.08$ is consistent with
the
$B \rightarrow
\pi^+ \pi^-$ measurement which gives $\Delta \alpha \sim 10^{\circ}$ (for
$\gamma \sim 90^{\circ}$).

\begin{figure}
\vspace{-0.5in}
\centerline{\epsfxsize=0.8\textwidth \epsfbox{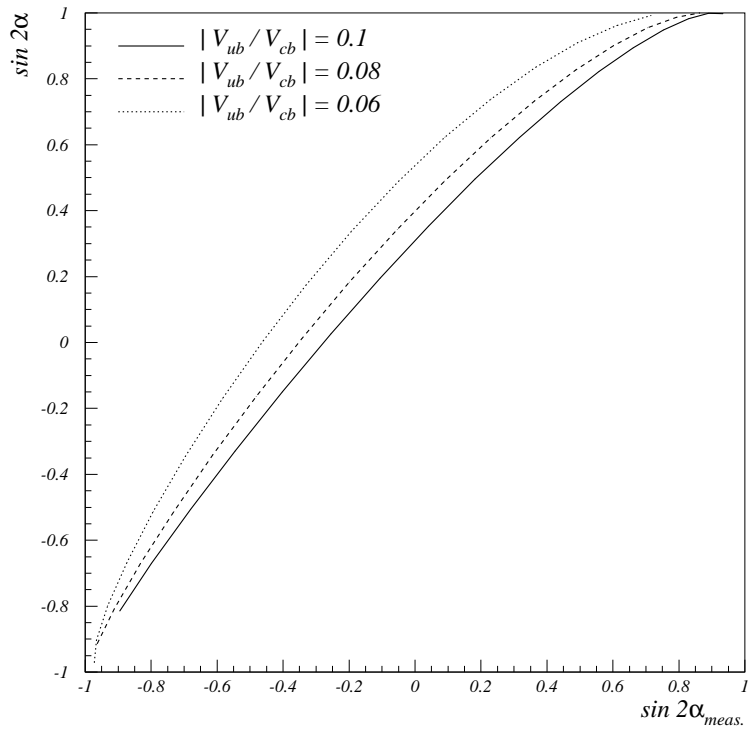}}
\vspace{-0.25in}
\caption{The ``true'' value of $\sin 2 \alpha$
as a function of the
value of $\sin 2 \alpha$ ``measured''
in $B \rightarrow \pi^+ \pi^-$ decays for $| V_{ub} / V_{cb} | = 0.1$
(solid curve), $0.08$ (dashed curve) and
$0.06$ (dotted curve). }
\protect\label{figrmalpha}
\end{figure}

The computation of Beneke
{\it et al.} \cite{beneke} includes
final state rescattering phases, {\it i.e.,} it is
{\em exact} up to $O( \Lambda _{QCD} / m_b)$ and $O (\alpha _s^2)$
corrections. Thus, the value of $\sin 2 \alpha$
``measured''
in $B \rightarrow \pi^+ \pi^-$ decays (Eq. (\ref{alphameas}))
is a known
function of $\gamma$ and $| V_{ub} / V_{cb} |$ only (in particular,
there is no dependence on the phenomenological parameter
$\xi \sim 1 / N$ and strong phases are included
unlike in the earlier factorization
framework \cite{ali}).
Since, the ``true'' value of $\alpha$
can also be expressed in terms of $\gamma$ and $| V_{ub} / V_{cb} |$
, we can estimate the ``true'' value of
$\sin 2 \alpha$
from the
``measured'' value of $\sin 2
\alpha$ for a given value of $| V_{ub} / V_{cb} |$
(of course, up to $O( \Lambda _{QCD} / m_b)$ and $O (\alpha _s^2)$
corrections);
this is shown in Fig. \ref{figrmalpha}
where we have restricted
$\gamma$ to be in the range $(40^{\circ}, 120^{\circ})$
as indicated by constraints on the
unitarity triangle from present data.
If $0^{\circ} \leq \gamma \leq 180^{\circ}$ is allowed, then there will
be a discrete ambiguity in the determination of
$\sin 2 \alpha$ from $\sin 2 \alpha _{meas.}$.

\section{Conclusions}
We have shown how $\alpha$ can be obtained from the measured value of
$\sin 2 \alpha$ inspite of large penguin effects. The theoretical work can be
extended to $K \pi$ modes to obtain values of $\gamma$ from
the measured branching ratios. Naive factorization suggest
$\gamma \approx 100 ^\circ$ \cite{he,dutta}.

\end{document}